\documentclass[conference]{IEEEtran}
\usepackage {cite}
\usepackage[]{graphicx}
\graphicspath{{./figures/}}
\usepackage[cmex10]{amsmath}
\usepackage[thmmarks,amsmath]{ntheorem}
\usepackage{accents}
\usepackage{ntheorem}
\usepackage{amssymb}
\usepackage{bbm}

\newtheorem{theorem}{Theorem}

\makeatletter
\def\widebar{\accentset{{\cc@style\underline{\mskip10mu}}}}
\def\Widebar{\accentset{{\cc@style\underline{\mskip8mu}}}}
\makeatother
\usepackage{amsmath,empheq} 
\usepackage{bm}
\usepackage{indentfirst}
\usepackage{algpseudocode}
\usepackage{algorithm}

\let\Phi\varPhi
\let\Omega\varOmega

\begin{document}
\IEEEoverridecommandlockouts

\title{Proactive Caching for Energy-Efficiency in Wireless Networks: A Markov Decision Process Approach}
\author{\IEEEauthorblockN{Zhijie Chen}
	\IEEEauthorblockA{Department of Electrical Engineering\\
		Stanford University\\
		Stanford, CA, USA, 94305\\
		Email: zcchen@stanford.edu}
	\and
\IEEEauthorblockN{Hoshyar Mohammed, and Wei Chen, \emph{Senior Member, IEEE}}
	\IEEEauthorblockA{Department of Electronic Engineering / TNList\\Tsinghua University\\ Beijing, China, 100084
		\\Email: mu-h16@mails.tsinghua.edu.cn, wchen@tsinghua.edu.cn\vspace{-2cm}}
	\thanks{This research was supported by the National Science Foundation of China under Grant Nos. 61671269 and 61322111, and the National 973 Program of China under Project No.2013CB336600.}}
\maketitle

\begin{abstract}
	
Content caching in wireless networks provides a substantial opportunity to trade off low cost memory storage with energy consumption, yet finding the optimal causal policy with low computational complexity remains a challenge. This paper models the Joint Pushing and Caching (JPC) problem as a Markov Decision Process (MDP) and provides a solution to determine the optimal randomized policy. A novel approach to decouple the influence from buffer occupancy and user requests is proposed to turn the high-dimensional optimization problem into three low-dimensional ones. Furthermore, a non-iterative algorithm to solve one of the sub-problems is presented, exploiting a structural property we found as \textit{generalized monotonicity}, and hence significantly reduces the computational complexity. The result attains close performance in comparison with theoretical bounds from non-practical policies, while benefiting from higher time efficiency than the unadapted MDP solution.

\end{abstract}


%

%


\IEEEpeerreviewmaketitle
\section{Introduction}

With the escalating growth of mobile data traffic caused by the proliferation of smart mobile devices, and the growing online services (i.e., Video on Demand (VoD) streaming, Facebook, Twitter,...), corresponding energy consumption is increasing considerably \cite{cisco}. Furthermore, video streaming and social applications require high bandwidth and strict delay constraint, which in turn has further adverse effects on user experience quality and Operational Expenses (OpEx). In addition, It has been reported that the network data traffic will continue growing in the future years \cite{cisco}. Hence, efficient energy utilization is of paramount importance in the design of future wireless networks. In this paper, our objective is to exploit the cache-enabled user devices to reduce the energy consumption by focusing particularly on the wireless transmission cost.

The massive mobile users generate issues to be addressed, such as, extremely high throughput and stringent Quality of Service (QoS) requirements, and excessive energy consumption. One way to deal with such issues is to deploy larger number of small cells and fog style access points \cite{6171992,6476878}. However, this approach does not address the energy efficiency and still large undesirable latency and network congestion could be induced during peak-traffic hours. Hence, content cashing at the edge of wireless networks has attracted much attention from both academia and industry \cite{cachinginwnetworks,latency,fundamental_caching_madah,maria16,contpushing_with_rdi_17}.

Moving contents to the edge of network emerged as a prospective technique, that can significantly reduce transmission latency\cite{cachinginwnetworks,latency}, control traffic load \cite{fundamental_caching_madah}, and reduce energy consumption \cite{maria16}, or can be employed to increase overall system throughput\cite{contpushing_with_rdi_17}. Content caching utilizes the recent advances in the field of context awareness and leverages on the low price memory storage to enhance the system performance. Content caching enables proactive transmission (i.e., \textit{pre-serve}) that allows the user to download content files over a longer period, hence, reducing energy consumption. It can also mitigate severe network condition  (i.e.,peak-traffic hours) by pushing at favorable transmission times. For instance, by exploiting the user demand information the base station (BS) can push the desired content files prior to the playback time (i.e., in video streaming). Hence, saving a significant amount of energy by avoiding the unfavorable channel conditions. It also provides another advantage by shifting traffic load (or equivalently reducing traffic variability) for better system utilization. In another line of research, the objective is to increase the throughput of content-centric wireless networks to better support the exponential growth of wireless data traffic. In \cite{contpushing_with_rdi_17} the available cache memory at the user is exploited via a joint pushing and caching method to increase the system throughput under non-causal, statistical, and causal knowledge of user request delay information (RDI). 

In this paper, we consider an optimization problem of a proactive caching wireless communications channel, with limited data buffer capacity available at the receiver end. Specifically, we formulate the joint pushing and caching (JPC) problem as an infinite horizon average cost Markov decision process (MDP) and devise a randomized policy to minimize the average energy consumption over time. In each timeslot the amount of data to transmit is a random variable, whose distribution is decided by the optimal policy, taking both the buffer occupancy and user requests into account. Numerically solving the optimization problem is computationally demanding due to the two-dimensional state space and randomized policy, known as the curse of dimensionality. Therefore, we develop a novel approach to decouple the factors of buffer occupancy and user requests by introducing the degenerated state space, and hence breakdown the optimization into three more tractable sub-problems. Furthermore, we found a structural property that gives a non-iterative way to design the optimal policy under certain constraints in the degenerated space. We name the property as \textit{generalized monotonicity}, which brings significant improvement to the computational complexity. The model in this work performs well in comparison against two theoretical bounds, which were derived from assuming non-causal knowledge of user request delay information (RDI)\cite{Zafer2009} and unlimited cache capacity, respectively.
\section{System Model}
\begin{figure}[htbp]
	\begin{center}
		\vspace{-0.5cm}
		\includegraphics[width=0.5\textwidth]{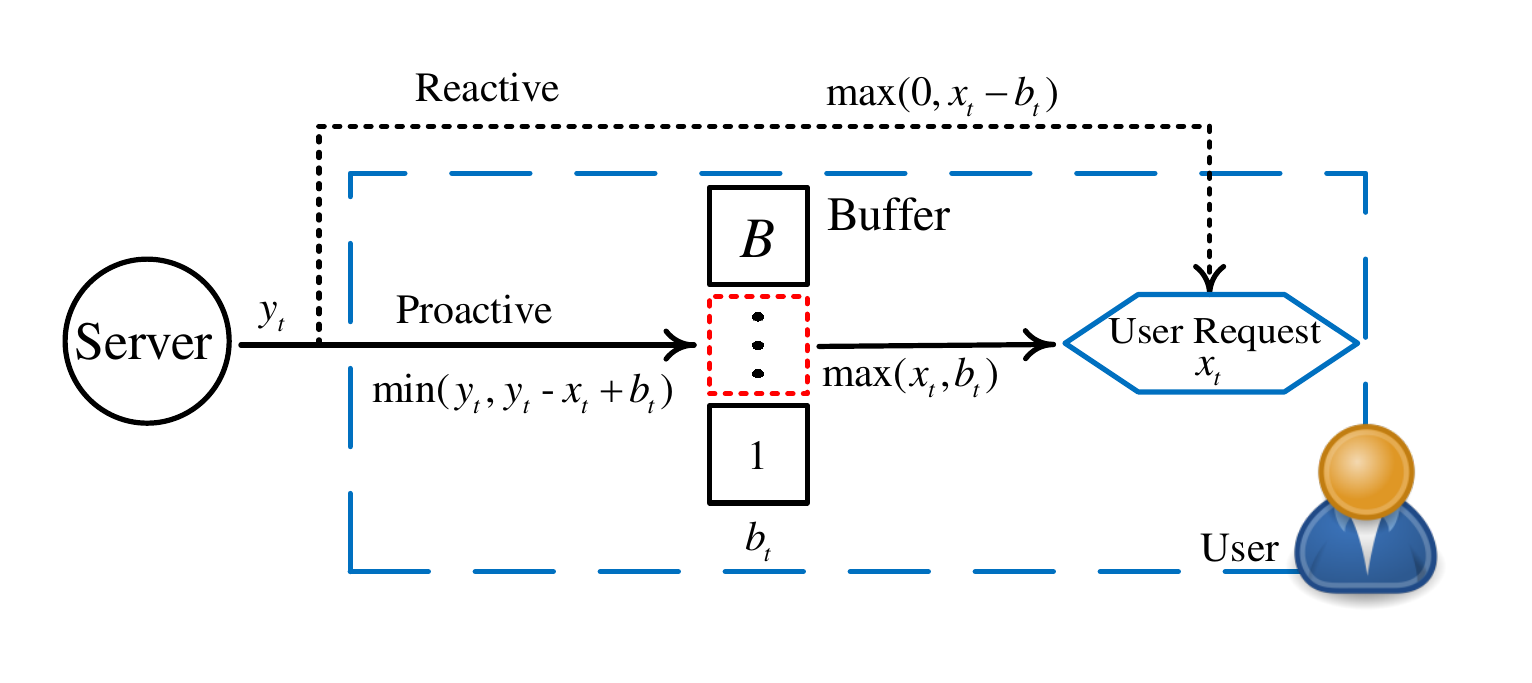}
		\vspace{-1cm}
		\caption[]{System model. As explained in section (\ref{sec2a}), fetching data from the buffer always has higher priority over via on-demand transmission, based on the assumption of temporal order. This observation gives the data flow on each edge.}
		\label{System_Model}
	\end{center}
\end{figure}

Consider a wireless communication system between a server and a user equipped with a limited buffer as depicted in Fig. \ref{System_Model}. The system operates over an infinite time horizon in a discrete time fashion, with timeslots $t=0,1,2,...$. At the beginning of each timeslot, the user requests for a certain amount of data, which must be fulfilled by the end of the timeslot. The data may be transmitted in advance, stored and read from the buffer (proactively), or be transmitted on-demand (reactively), or combined. Also, we consider the scenario where data stream has been temporally ordered, i.e., the server knows what to be requested in the near future, while not knowing how soon the requests will be made.
\subsection{Request model and buffer state}\label{sec2a}
We assume that data is requested in content items of identical size. Thus, we may denote the number of content items requested in timeslot t as $x_t \in\mathcal{X}\triangleq\{0,1,\dots,X\}$, where $X$ is the maximum possible value of $x_t$. We further assume that $x_t, t=0,1,\dots$ are \textit{i.i.d.} integer random variables bounded by $X$, yielding probability mass function (p.m.f) $f_X(\cdot)$. One can always limit $x_t$ to integer values by choosing a small enough size for content items. Let a $(X+1)$-dimensional vector $\bm{p}$ denotes the p.m.f. of $x_t$, i.e., $p_x=f_X(x), x\in\mathcal{X}$.

Let buffer state $b_t$ denote the number of content items at the beginning of timeslot $t$, $b_t\in\mathcal{B}\triangleq\{0, 1, \dots,B\}$.This implies the buffer storage capacity being $B$ content items. Since data stream is well-ordered, for any two content items, we always know which one would be requested first, and thus which to proactively push first. Therefore, as long long as the buffer is not empty, there is no reason to transmit data on-demand. In other words, on-demand transmission happens if and only if $x_t>b_t$.
%
\subsection{System State Model}
	The system state space is denoted by $\mathcal{S}$ which consists of $b$ and $x$ elements and has the cardinality of $(B+1)\times(X+1)$. The pair $(b_t,x_t)$ constitutes each state $s_t\in \mathcal{S}$. Furthermore, we define degenerated state $\mathfrak{b}_b=\{s: s=(b,x), x\in\mathcal{X}\}$, and degenerated state space $\mathfrak{B}=\{\mathfrak{b}_b: b\in\mathcal{B}\}$. When it does not cause confusion, we use $\mathfrak{b}_b$ and $b$ interchangeably. 

\begin{figure}[htbp]
	\begin{center}
		\vspace{-0.6cm}
		\includegraphics[width=0.4\textwidth]{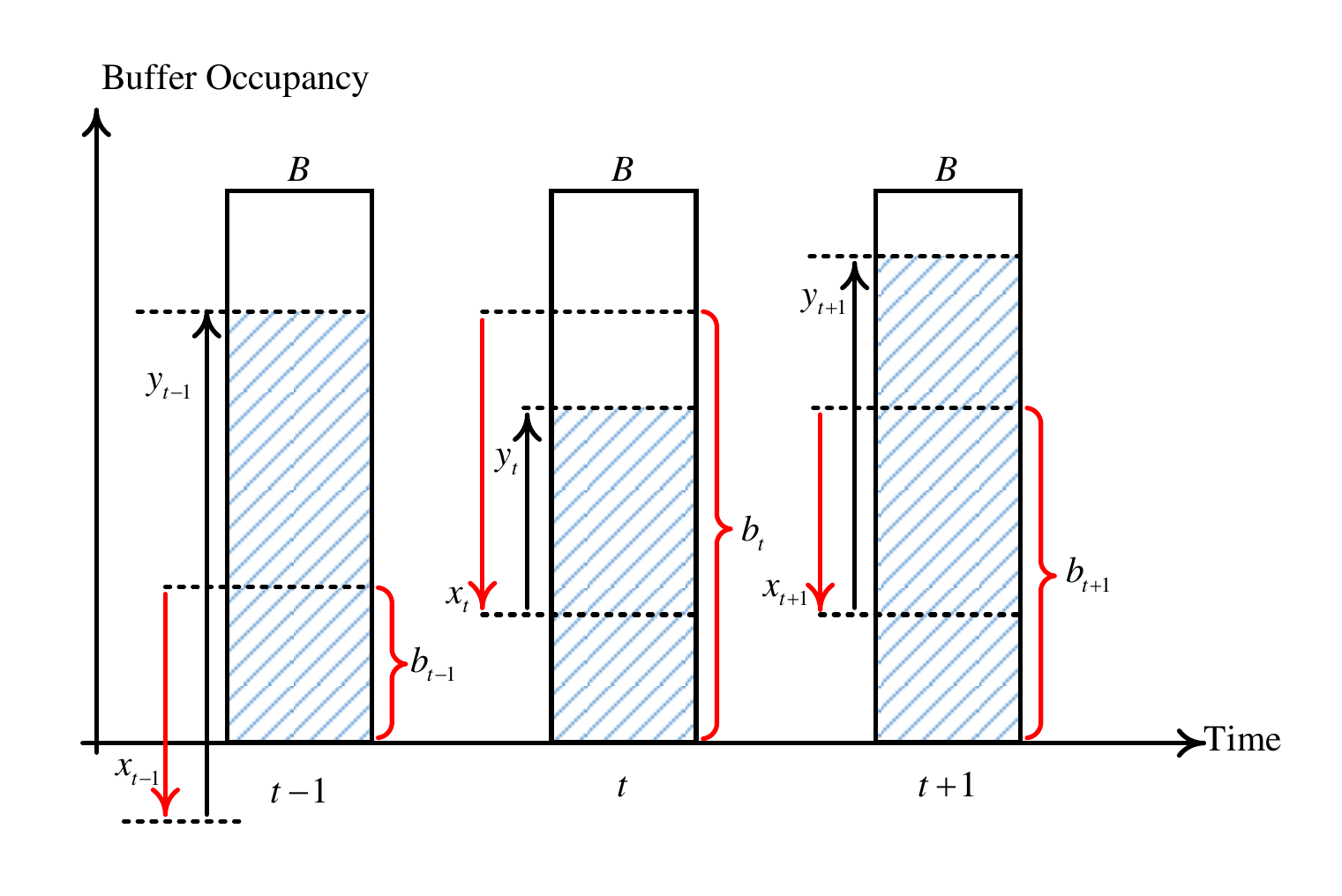}
		\vspace{-0.6cm}
		\caption[]{Buffer state evolution. An illustration of the relationship between the buffer occupancy $b_t$, user requests $x_t$, and data transmission $y_t$ in a timeslot.}
		\label{cache_state}
	\end{center}
\end{figure}

Let $y_t$ denote the number of content items downloaded within timeslot $t$. Under a given policy, $y_t$ is a random variable with distribution being contingent on buffer occupancy $b_t$ and user requests $x_t$, i.e., $s_t$. Buffer occupancy evolves as a time-homogeneous Markov chain which can be described as
\begin{equation}
b_{t+1}=b_t+y_t-x_t,
\label{bt_evolve}
\end{equation}
where
\begin{equation}\begin{aligned}
0\leq b_t\leq B, && \forall t.
\end{aligned}
\label{bt_constraint}
\end{equation}
the first inequality in (\ref{bt_constraint}) comes from the model setting that requests must be fulfilled within the requested timeslot $t$, and thus $b_t+y_t\geq x_t$. The Fig. \ref{cache_state} depicts the dynamical relation between $y_t$, $b_t$ and $x_t$ in each timeslot. A policy $\pi$ would determine the distribution of $y_t$ from $b_t$ and $x_t$.

We consider a causal knowledge of user RDI, that the user requests for a future timeslot stays unknown until the beginning of the timeslot.

The transmission action in each timeslot incurs a certain energy cost, whose corresponding power is typically a convex and continuous function of data rate. Conventionally we assume this function to be exponential\cite{gungurgunduz}. By \cite{Zafer2009}, constant transmission rate should be used within a timeslot to minimize energy consumption. Absorbing all constants and normalizing the scalar in our model, we give the energy consumption in timeslot $t$ by:
\begin{equation}
\rho_t=\rho(y_t)=\eta^{y_t}-1,
\label{eq3}
\end{equation} where $\rho_t$ is the energy consumption and $\eta>1$ is a constant. Since, $y_t$ takes finitely many values, it can be shown that $\mathbf{E}\rho_t<\infty$.
%
\section{A Randomized MDP problem Formulation}
\begin{figure}[h]
	\begin{center}
		\vspace{-.5cm}
		\includegraphics[width=0.5\textwidth]{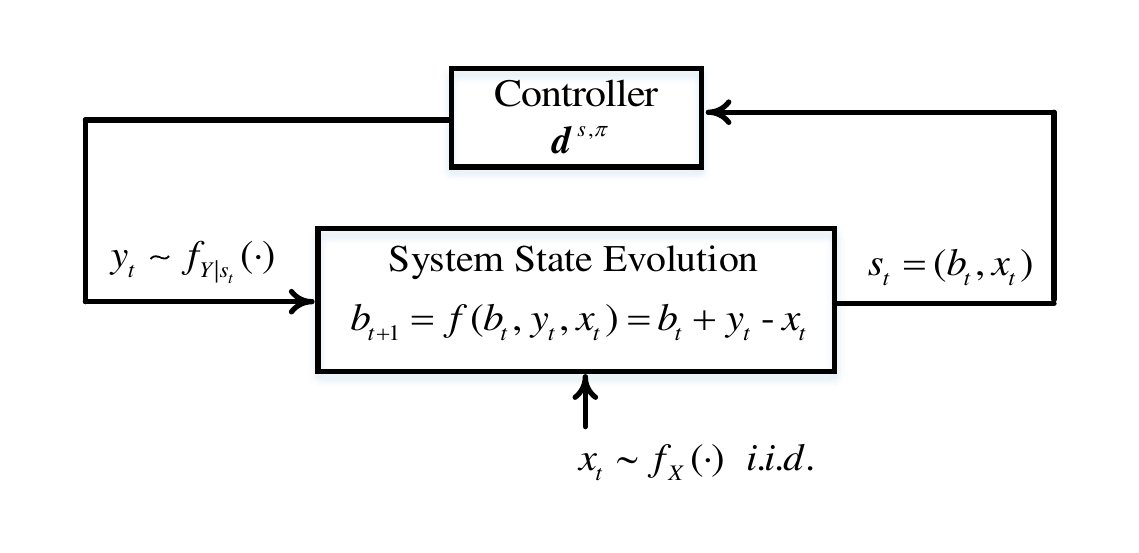}
		\vspace{-1cm}
		\caption[]{At each timeslot $t$ the controller, i.e., the server, observes the current system state $s_t$ and applies a control $y_t\sim f_{Y\mid s_t}(\cdot)$ (and equivalently, $b_{t+1}\sim\bm{d}^{s_t,\pi}$) contingent on the state.}%
		\label{policy1}%
	\end{center}%
\end{figure}

In a randomized policy, a transmission action $y_t$ yields a distribution which we design contingent on $s_t$, i.e., $y_t \sim f_{Y|s_t}(y)$. The optimal transmission policy selects a distribution for $y_t$. However, the range where random variable $y_t$ takes non-zero probability changes given different $s_t$. By (\ref{bt_evolve}), $\sigma(y_t\mid s_t) = \sigma(b_{t+1}|s_t)$. Therefore, we study the conditional probability of $b_{t+1}$ instead of $y_t$. The system evolves as depicted in Fig. \ref{policy1}.

Let a $(B+1)$-dimensional vector $\bm{d}^{s,\pi}$ denote a randomized decision, i.e., p.m.f., conditioning on state $s$ under policy $\pi$, where $d^{s,\pi}_{b^+}\triangleq\Pr\left(b_{t+1}={b^+}\mid s_t=s,\pi\right)$. All $(\bm{d}^{s,\pi})_{s\in\mathcal{S}}$ give the transition probabilities in $\mathcal{S}$ and $\mathfrak{B}$. Denote the transition probability matrix in $\mathfrak{B}$ under policy $\pi$ by $\bm{A}^\pi\triangleq{[\bm{a}^{0,\pi},\bm{a}^{1,\pi},...,\bm{a}^{B,\pi}]}^\intercal$, where $\bm{a}^{b,\pi}=\sum_{x}\bm{d}^{(b,x),\pi}p_x$.
	Since every state in $\mathcal{S}$ has to be assigned with a randomized decision, the corresponding policy space is $(B+1)^2\times(X+1)$- dimensional. Denote the probability that the system occupies state $(b,x)\in\mathcal{S}$ in timeslot $t$ under policy $\pi$ as $q_{b,x}^{t,\pi}\triangleq\Pr\left(s_t=(b,x)\mid\pi\right)$, and define matrix $\bm{Q}^{t,\pi}\triangleq{(q^{t,\pi}_{b,x})}_{b\in\mathcal{B},x\in\mathcal{X}}$, which contains the probabilities for all states. Similarly, we denote the state probability of $\mathfrak{b}_b\in\mathfrak{B}$ in timeslot $t$ under policy $\pi$ as $r_b^{t,\pi}\triangleq\Pr\{b_t=b\mid\pi\}$, and define vector $\bm{r}^{t,\pi}\triangleq(r_b^{t,\pi})_{b\in\mathcal{B}}$. Since, $x_t\sim f_X(\cdot)$ is i.i.d., we have 
\begin{equation}
\bm{Q}^{t,\pi}=\bm{r}^{t,\pi}{\bm{p}}^\intercal,\hspace{0.5cm} \forall t>0, \pi.
\end{equation}
Under a given policy $\pi$, the expected energy cost in timeslot $t$ solely depends on state $s_t$. We define the expected energy cost for state $s=(b,x)$ as $\omega_s^\pi\triangleq \mathbf{E}[\rho_t|s_t=s,\pi]$. Let $\bm{\Phi}^B\triangleq{[1,\eta,\eta^2,\dots,\eta^B]}^\intercal$ and $\bm{\Phi}^X\triangleq{[1,\eta,\eta^2,\dots,\eta^X]}^\intercal$, we have
\begin{equation}
\omega_s^\pi=\eta^{x-b}\sum_{m=0}^{B}\eta^{m}d_m^{s,\pi}-1=\eta^{x-b}{\bm{\Phi}^B}^\intercal\bm{d}^{s,\pi}-1.
\end{equation}
denote with matrix $\bm{\Omega}^\pi\triangleq{(\omega^\pi_{b,x})}_{b\in\mathcal{B},x\in\mathcal{X}}$ the expected energy costs for all states. Under the average reward criterion, we formally state the optimization problem as
\begin{subequations}
	\begin{align}
	& \underset{\pi}{\text{minimize}}
	& & \mathcal{L}(\pi)={\bm{r}^{\infty,\pi}}^\intercal\bm{\Omega}^\pi\bm{p}\label{prb6a}\\
	& \text{subject to} & & {\bm{A}^\pi}^\intercal\bm{r}^{\infty,\pi}=\bm{r}^{\infty,\pi},\label{prb6b}\\
	& & &  \bm{d}^{s,\pi}\geq 0,& \forall s,\label{prb6c}\\
	& & & \bm{1}^\intercal\bm{d}^{s,\pi}=1,&  \forall s,\label{prb6d}\\
	& & & d_m^{(b,x),\pi}=0,& \forall m <b-x.\label{prb6e}
	\end{align}
	\label{opt_1}%
\end{subequations}
where constraint (\ref{prb6b}) implies stationary distribution and (\ref{prb6c}) and (\ref{prb6d}) ensure that $\bm{d}^{s,\pi}$ corresponds to a p.m.f.. The last constraint rules out the possibility of $y_t$ being negative, since the number of content item transmitted must be positive or $0$.

As the policy space is $(B+1)^2\times(X+1)$-dimensional, problem (\ref{opt_1}) is an optimization with high dimensionality, whose convexity is hard to determine. Value iteration being a popular algorithm regarding solving MDP problems, overwhelming computational complexity is yet a problem if to run value iteration in $\mathcal{S}$. We found a novel approach to decouple the influence from $b_t$ and $x_t$ and hence turn (\ref{opt_1}) to three sub-problems, which are low-dimensional and more tractable. By solving the three sub-problems specified below, we equivalently solves (\ref{opt_1}).

First, we assume a transition matrix in $\mathfrak{B}$ is given as $\bm{A}$ and find the optimal policy, i.e., $\{\bm{d}^{s,\pi}: s\in\mathcal{S}\}$, among all policies that results in $\bm{A}$. Secondly, we find the optimal $\bm{A}$ whose corresponding optimal policy minimizes the expected energy consumption in one iteration step. Thirdly, we carry out value iteration in $\mathfrak{B}$, each step searching for the optimal $\bm{A}$ iteratively, until the transition matrix converges to $\bm{A}^*$. By \cite{puterman2014markov}, $\bm{A}^*$ is the minimizer for average energy cost among all possible $\bm{A}$. We argue that we have also found the optimal $\pi^*$, among all policy space, that minimize the average energy cost as
\begin{equation}
\underset{\bm{A}}{\min}\left(\underset{\pi}{\min}\mathcal{L}(\pi)\mid\bm{A}\right)=\underset{\pi}{\min}\mathcal{L}(\pi).
\end{equation}
We next elaborate each of the three sub-problems.
\subsection{Optimal Decisions with Given Transition Matrix in $\mathfrak{B}$}
	\label{sectionIIIb}
	Denote with $\omega_b^\pi$ the expected energy cost under policy $\pi$ in a timeslot belonging to state $\mathfrak{b}_b$. We have
	\begin{equation}
	\omega_b^\pi\triangleq\mathbf{E}[\rho_t|b_t=b,\pi]=\sum_{x=0}^{X}p_x\omega^\pi_{(b,x)}=\bm{\Omega}^\pi_{(b,:)}\bm{p},
	\label{exp_enrg_cost}
	\end{equation}
	where $\bm{\Omega}_{(b,:)}$ denotes the row vector from the $b^{th}$ row of $\Omega$.
	Recall that $\bm{A}^\pi={\left[\begin{array}{cccc}
		\bm{a}^{0,\pi}&\bm{a}^{1,\pi}&\dots&\bm{a}^{B,\pi}
		\end{array}\right]}^\intercal$ and $\bm{a}^{b,\pi}=\sum_{x}\bm{d}^{(b,x),\pi}$, i.e.,
	the $b^{th}$ row of $\bm{A}$ is solely determined by decisions from $\mathfrak{b}_b$. We formulate a matrix that represents decisions from $\mathfrak{b}_b$ as $\bm{D}^{b,\pi}\triangleq{\left[\begin{array}{cccc}
		\bm{d}^{(b,0),\pi}&\bm{d}^{(b,1),\pi}&\dots&\bm{d}^{(b,X),\pi}
		\end{array}\right]}^\intercal$. The first sub-problem, finding the optimal decisions under the constraint of a fixed $\bm{A}$, is formally stated as:
	\begin{subequations}
		\renewcommand{\theequation}{\theparentequation.\alph{equation}}
		\begin{align}
		& \underset{\bm{D}^{b,\pi}}{\text{minimize}}
		& & \omega_b^\pi=\eta^{-b}{\bm{\Phi}^X}^\intercal\mathrm{diag}(\bm{p})\bm{D}^{b,\pi}\bm{\Phi}^B-1\label{prb5a}\\
		& \text{subject to} & & {\bm{D}^{b,\pi}}^\intercal\bm{p}=\textbf{\textit{a}}^\pi_b,\label{prb5b}\\
		& & &  \bm{D}^{b,\pi}\bm{1}=\bm{1},\label{prb5c}\\
		& & & D^{b,\pi}_{m,n}\geq 0,\hspace{1cm}  \forall m,n\label{prb5d}\\
		& & & D^{b,\pi}_{m,n}=0, \hspace{1cm} \forall m+n<b\label{prb5e},
		\end{align}%
		\label{opt_2}%
	\end{subequations}%
	for every $b\in\mathcal{B}$. We define a function $h: \mathbb{R}^{X+1}\rightarrow \mathbb{R}$ as:
	\begin{equation}
	h(\bm{a}^{b,\pi})=\min_\pi\left( \omega^\pi_b|{\bm{a}^{b,\pi}}\right),
	\end{equation} in light of the optimization problem (\ref{opt_2}). In section \ref{sectionIV}, we propose an efficient algorithm to find $h(\bm{a}^\pi_b)$ without iteration, exploiting a certain structural property we name as \textit{generalized monotonicity}.

\begin{subsection}{Optimal Transition Probability Matrix in $\mathfrak{B}$}
	\label{subprob2}
	In a finite MDP with $N$ timeslots, define $\upsilon^t_b$ as the expected total cost over period $\{N-t+1, N-t+2, ..., N\}$, starting from $b_{N-t+1}=b$. Denote vector $\bm{\upsilon}^t\triangleq(\upsilon^t_b)_{b\in\mathcal{B}}$. The iteration from $t-1$ to $t$ can be written as
	\begin{subequations}
		\renewcommand{\theequation}{\theparentequation.\alph{equation}}
		\begin{align}
		& \underset{\bm{a}^{b,\pi}}{\text{minimize}}
		& & \upsilon^t_b=h(\bm{a}^{b,\pi})+{\bm{a}^{b,\pi}}^\intercal\bm{\upsilon}^{t-1}\label{eq.13a}\\
		& \text{subject to} & & \bm{a}^{b,\pi}\geq0,\label{eq.13b}\\
		& & & \bm{1}^\intercal\bm{a}^{b,\pi}=1,\label{eq.13c}\\
		& & & a^{b,\pi}_m\leq\sum_{x=b-m}^{X}p_x,\quad \forall m=0,1,...,b.\label{eq.13d}
		\end{align}%
		\label{opt_3}%
	\end{subequations}%
(\ref{eq.13b}) and (\ref{eq.13c}) are the nature of p.m.f., and (\ref{eq.13d}) is derived from (\ref{prb5b}) and (\ref{prb5e}).
Now we prove that optimization problem (\ref{opt_3}) is convex, and hence can be easily solved by conventional optimization tools, say, the interior-point method.
	
	\begin{IEEEproof}
		Let $\bm{a}^{b,\pi_1}$ and $\bm{a}^{b,\pi_2}$
		be two feasible points of (\ref{opt_3}), whose corresponding solutions in (\ref{opt_2}) are $\bm{D}^{b,\pi_1}$ and $\bm{D}^{b,\pi_2}$, respectively. For all $\lambda\in\{\lambda\in\mathbb{R}|0<\lambda<1\}$, let
		\begin{equation}
		\bm{a}^{b,\pi'}=(1-\lambda)\bm{a}^{b,\pi_1}+\lambda\bm{a}^{b,\pi_2},
		\end{equation}  and, let 
		\begin{equation}
		\bm{D}^{b,\pi'}=(1-\lambda)\bm{D}^{b,\pi_1}+\lambda\bm{D}^{b,\pi_2}.
		\end{equation}
		it can be directly shown that $\bm{a}^{b,\pi'}$ and $D^{b,\pi'}$ satisfy constraint sets (\ref{eq.13b})-(\ref{eq.13d}) and (\ref{prb5c})-(\ref{prb5e}) respectively. Also we have
		\begin{equation}
		\begin{aligned}
		{\bm{D}^{b,\pi'}}^\intercal\bm{p}=& ((1-\lambda)\bm{D}^{b,\pi_1}+\lambda\bm{D}^{b,\pi_2})^\intercal\bm{p}\\
		=& (1-\lambda)\bm{a}^{b,\pi_1}+\lambda\bm{a}^{b,\pi_2}\\
		=& \bm{a}^{b,\pi'},
		\end{aligned}
		\end{equation}%
		i.e., $\bm{D}^{b,\pi'}$ satisfies (\ref{prb5b}) with $\bm{a}^{b,\pi'}$. Thus, $\bm{a}^{b,\pi'}$ and $\bm{D}^{b,\pi'}$ are feasible points of (\ref{opt_3}) and (\ref{opt_2}) respectively. Now denote with $\omega_b^{\pi'}$ the expected average energy cost carried out by $\bm{D}^{b,\pi'}$. Because (\ref{prb5a}) is linear on $\bm{D}^{b,\pi}$, $\omega_b^{\pi'}=(1-\lambda)h(\bm{a}^{b,\pi_1})+\lambda h(\bm{a}^{b,\pi_2})$, which gives a feasible value of (\ref{opt_2}) under $\bm{a}^{b,\pi'}$.
		By definition,
		\begin{equation}
		\begin{aligned}
		h(\bm{a}^{b,\pi'})=&\min_\pi\left(\omega^\pi_b\mid\bm{a}^{b,\pi'}\right)\\
		\leq& \omega^{\pi'}_b\\
		=& (1-\lambda)h(\bm{a}^{b,\pi_1})+\lambda h(\bm{a}^{b,\pi_2}).
		\end{aligned}
		\end{equation}
		thus, $h$ is convex. Since that ${\bm{a}^{b,\pi}}^\intercal\bm{\upsilon}^{t-1}$ is linear, optimization problem (\ref{opt_3}) is convex.
	\end{IEEEproof}
\end{subsection}
\begin{subsection}{Value Iteration in degenerated state space $\mathfrak{B}$}
	
	By \ref{sectionIIIb} and \ref{subprob2}, we find the optimal transition matrix in $\mathfrak{B}$ and its corresponding policy, in light of one-step iteration. Now we start from:\begin{equation}
	\bm{\upsilon}^0=\bm{0},
	\end{equation}
	and carry out the optimality equations given by
	\begin{equation}
	\begin{aligned}
	\upsilon^t_b=& \min_{\bm{a}^{b,\pi}} h(\bm{a}^{b,\pi})+\sum_{b^+=0}^Ba^{b,\pi}_{b^+}\upsilon^{t-1}_{b^+}\\
	=& \min_{\bm{a}^{b,\pi}} h(\bm{a}^{b,\pi})+{\bm{a}^{b,\pi}}^\intercal\bm{\upsilon}^{t-1},
	\end{aligned}
	\label{optimality_equs}
	\end{equation} for $\forall b\in\mathcal{B}$ and $t=1,2,\dots$. The transition matrix $A^\pi$ in $\mathfrak{B}$ and its corresponding policy $\pi$ is updated on each iteration step. By \cite{puterman2014markov}, the iteration would eventually
	converge to satisfy the $\varepsilon$-optimal stopping criterion:
	\begin{equation}
	\label{convergence}
	\left|\max_i(\upsilon^t_i)-\min_j(\upsilon^t_j)\right|<\varepsilon.
	\end{equation}
	when (\ref{convergence}) holds, the transition matrix $A^{\pi*}$ in $\mathfrak{B}$ and its corresponding $\pi^*$ give a global $\varepsilon$-optimal policy. By taking limit $\varepsilon\rightarrow0$, the global optimum of (\ref{opt_1}) is attained.
\end{subsection}

\section{Generalized Monotonic Structure}
\label{sectionIV}
As mentioned in subsection \ref{sectionIIIb}, an efficient algorithm that solves problem (\ref{opt_2}) without iteration is proposed in this section, and its corresponding time complexity analysis is given in section \ref{sectionV}. A special pattern of the optimal decision matrix $D^{b,\pi}$ is revealed by theorem 1, which we name \textit{generalized monotonicity}.
\subsection{Solution of Optimization without Iteration} 

Intuitively the \textit{generalized monotonicity} captures the feature that the optimal decision matrix has all its non-zero entries lying in a stripe expanding from the top-right corner to the bottom-left. If a entry is non-zero, then the block adjacent to its bottom-right corner shold be all-zero. We formally describe and prove it as the following theorem:
\begin{theorem}
	\label{theorem1}
	If $\bm{D}^{b,\pi^*}$ is a solution of (\ref{opt_2}), and there exist $ x^-, b^-$ s.t. $D^{b,\pi^*}_{x^-,b^-}>0$, then $D^{b,\pi^*}_{x^+,b^+}=0$ for $\forall x^+\in\{x\in\mathbb{N}: x^-<x\leq X\}, b^+\in\{b\in\mathbb{N}: b^-<b\leq B\}$.
\end{theorem}

\begin{IEEEproof}
	Suppose there exist $x^-, b^-$ s.t. $0\leq x^-<X, 0\leq b^-< B$ and $D^{b,\pi^*}_{x^-,b^-}>0$ (if not, the case is trivial and Theorem \ref{theorem1} still holds). We prove by contradiction, starting by assuming that $\exists x^+, b^+\in\mathbb{N}$ s.t. $x^-<x^+\leq X$, $b^-<b^+\leq B$ and $D^{b,\pi^*}_{x^+,b^+}>0$. Now we pick any positive number $\delta$ that satisfies $p_{b^+}\delta<D^{b,\pi^*}_{x^-,b^-}$ and $p_{b^-}\delta<D^{b,\pi^*}_{x^+,b^+}$. We formulate another decision matrix $D^{b,\pi'}$ which is identical to $D^{b,\pi^*}$ except the following elements
	\begin{subequations}
		\renewcommand{\theequation}{\theparentequation.\alph{equation}}
		\begin{align}
		&D^{b,\pi'}_{x^-,b^-}=D^{b,\pi^*}_{x^-,b^-}-p_{b^+}\delta\\
		&D^{b,\pi'}_{x^-,b^+}=D^{b,\pi^*}_{x^-,b^+}+p_{b^+}\delta\\
		&D^{b,\pi'}_{x^+,b^+}=D^{b,\pi^*}_{x^+,b^+}-p_{b^-}\delta\\
		&D^{b,\pi'}_{x^+,b^-}=D^{b,\pi^*}_{x^+,b^-}+p_{b^-}\delta
		\end{align}
	\end{subequations}
	because $\bm{D}^{b,\pi^*}$ is a solution of (\ref{opt_2}), it is easy to verify that $\bm{D}^{b,\pi'}$ satisfies (\ref{prb5b}) and (\ref{prb5c}). Also, (\ref{prb5d}) and (\ref{prb5e}) are naturally satisfied by the choice of $\delta$. Hence, $D^{b,\pi'}$ is a feasible point of (\ref{opt_2}). Further we have
	\begin{equation}
	\begin{aligned}
	&\quad\omega^{\pi^*}_b-\omega^{\pi'}_b\\
	&=\eta^{-b}{\bm{\Phi}^X}^\intercal\mathrm{diag}(\bm{p})(\bm{D}^{b,\pi^*}-\bm{D}^{b,\pi'})\bm{\Phi}^B\\
	&=\eta^{-b+x^-+b^-}p_{x^+}p_{x^-}(\eta^{x^+-x^-}-1)(\eta^{b^+-b^-}-1)>0.
	\end{aligned}
	\end{equation}
$\bm{D}^{b,\pi'}$ results in a smaller objective value in (\ref{prb5a}) than $D^{b,\pi^*}$, which contradicts the prerequisite that $D^{b,\pi^*}$ is a solution. Therefore, the original assumption must be false, proving the theorem.
\end{IEEEproof}
Now we formally state the Fast Assignment of State Transition (FAST) algorithm that gives a solution to the problem in (\ref{opt_2}) without iteration.
\begin{algorithm}
	\caption{The FAST Algorithm}
	
	\begin{algorithmic}[1]
		\State Initialization: $m=0, n=B, u_0=p_0, w_B=a^{b,\pi}_B, \bm{D}^{b,\pi}=\bm{0}$; 
		\If{ $u_m<w_n$, }\label{goto2}
		\State $D^{b,\pi}_{m,n}\triangleq\frac{u_m}{p_m}, w_n\triangleq w_n-u_m, m\triangleq m+1, u_m\triangleq p_m$, go to \ref{goto9};
		\EndIf
		\If{$u_m>w_n$,}
		\State  $D^{b,\pi}_{m,n}\triangleq\frac{w_n}{p_m}, u_m\triangleq u_m-w_n, n\triangleq n-1, w_n\triangleq a^{b,\pi}_n$, go to \ref{goto9};
		\EndIf
		\State $D^{b,\pi}_{m,n}\triangleq\frac{w_n}{p_m}, m\triangleq m+1, n\triangleq n-1, u_m\triangleq p_m, w_n\triangleq a^{b,\pi}_n$;
		
		\If{$m\leq X$ and $n\geq 0$,}\label{goto9}
		\State  go to \ref{goto2};
		\EndIf
		\State $\bm{D}^{b,\pi^*}\triangleq \bm{D}^{b,\pi}$;
	\end{algorithmic}
\end{algorithm}
Validity of the FAST algorithm is given by the following Theorem
\ref{theorem2}.
\begin{theorem}
\label{theorem2}
Let $\bm{D}^{b,\pi^*}$ be a solution of (\ref{opt_2}). $\forall m,n\in\mathbb{N}$ s.t. $0\leq m\leq X, 0 \leq n\leq B$, if denote 
\begin{subequations}
	\renewcommand{\theequation}{\theparentequation.\alph{equation}}
	\begin{align}
	&u_m=(a^{b,\pi}_n-\sum_{i=0}^{m-1}D^{b,\pi^*}_{i,n}p_i)/p_m\\
	&w_n=1-\sum_{i=n+1}^BD^{b,\pi^*}_{m,i},
	\end{align}
\end{subequations}
then $D^{b,\pi^*}_{m,n}=\min(u_m, w_n)$.
\end{theorem}
\begin{IEEEproof}
	Let $m,n\in\mathbb{N}$ s.t. $0\leq m\leq X, 0 \leq n\leq B$. By (\ref{prb5b}) to (\ref{prb5d}), we have
	\begin{subequations}
		\renewcommand{\theequation}{\theparentequation.\alph{equation}}
		\begin{align}
		&\sum_{i=0}^{m}D^{b,\pi^*}_{i,n}p_i\leq a^{b,\pi}_n
		\label{eq22a}\\
		&\sum_{i=n}^BD^{b,\pi^*}_{m,i}\leq 1
		\label{eq22b}
		\end{align}
	\end{subequations}
	the proof is given by contradiction. Assume that neither equation in (\ref{eq22a}) and (\ref{eq22b}) holds. By (\ref{prb5b}) $\sum_{i=0}^{X}D^{b,\pi^*}_{i,n}p_i=a^{b,\pi}_n,$ and also we have $	\sum_{i=0}^{m}D^{b\pi^*}_{i,n}p_i<a^{b,\pi}_n,$ Therefore
	\begin{equation*}
	\exists m^+>m \hspace{1cm}\text{s.t.}\hspace{1cm} D^{b,\pi^*}_{m^+,n}>0,
	\end{equation*}
by (\ref{prb5c}) and the assumption that the equation in (\ref{eq22b}) does not hold, we have $\exists n^-<n$ s.t. $D^{b,\pi^*}_{m,n^-}>0$. Therefore, we have $D^{b,\pi^*}_{m,n^-}>0$ and $D^{b,\pi^*}_{m^+,n}>0$, contradicting Theorem \ref{theorem1}. Thus, the original assumption is false, i.e., at least one of the equations in (\ref{eq22a})(\ref{eq22b}) holds. Rewrite (\ref{eq22a}) and (\ref{eq22b}) as
	\begin{subequations}
		\renewcommand{\theequation}{\theparentequation.\alph{equation}}
		\begin{align}
		&D^{b,\pi^*}_{m,n}\leq (a^{b,\pi}_n-\sum_{i=0}^{m-1}D^{b,\pi^*}_{i,n}p_i)/p_m\\
		&D^{b,\pi^*}_{m,n}\leq 1-\sum_{i=n+1}^BD^{b,\pi^*}_{m,i},
		\end{align}
	\end{subequations}
	it is clear that the equation with smaller value on the right side should hold, hence proving theorem \ref{theorem2}.
\end{IEEEproof}

\subsection{Complexity Analysis}
\label{sectionV}
Suppose the required precision of value iteration is $\epsilon_v$, and interior-point method is applied for all convex optimizations. By\cite{puterman2014markov}, the number of iteration steps is bounded by $-C_1\log\epsilon_v$, where $C_1$ is a constant. $B+1$ convex optimizations are solved with interior-point method in each iteration step. If precision $\epsilon_i$ is required for interior-point method, the number of times Newton's Method is called in each interior-point method is bounded by $-\sqrt{2(B+1)}\log\epsilon_i$ \cite{boyd2004convex}. In Newton's Method the FAST algorithm, whose time complexity is $O(X+B)$, is called $B+1$ times. Therefore, the time cost of value iteration in $\mathfrak{B}$ with the FAST algorithm is bounded by $O((X+B)B^{2.5}\log\epsilon_i\log\epsilon_v)$.

With similar analysis, we have the time complexity of value iteration in $\mathcal{S}$ which is $O(X^2B^{3.5}\log\epsilon_i\log\epsilon_v)$. An interesting insight into the two complexities is that when $X\ll B$ the difference between value iterations in $\mathfrak{B}$ and $\mathcal{S}$ fades out, which is the case that buffer size is much larger than expected data request in a timeslot. Since there is nearly unlimited caching space, an optimal policy becomes meaningless. However, in most cases, value iteration in $\mathfrak{B}$ with the FAST algorithm brings significant improvement to time efficiency.

\section{Simulation Results}
We first demonstrate the reasonably small compromise of the causal MDP method proposed in this paper, in comparison with the non-causal \textit{tightest string method} which attains the absolute optimum \cite{Zafer2009}. Then we compare the time consumption of value iteration in $\mathfrak{B}$, as proposed in this paper, and the conventional MDP method of value iteration in $\mathcal{S}$.
\begin{figure}[htbp]
	\centering
	\vspace{-0.4cm}
	\includegraphics[width=3in]{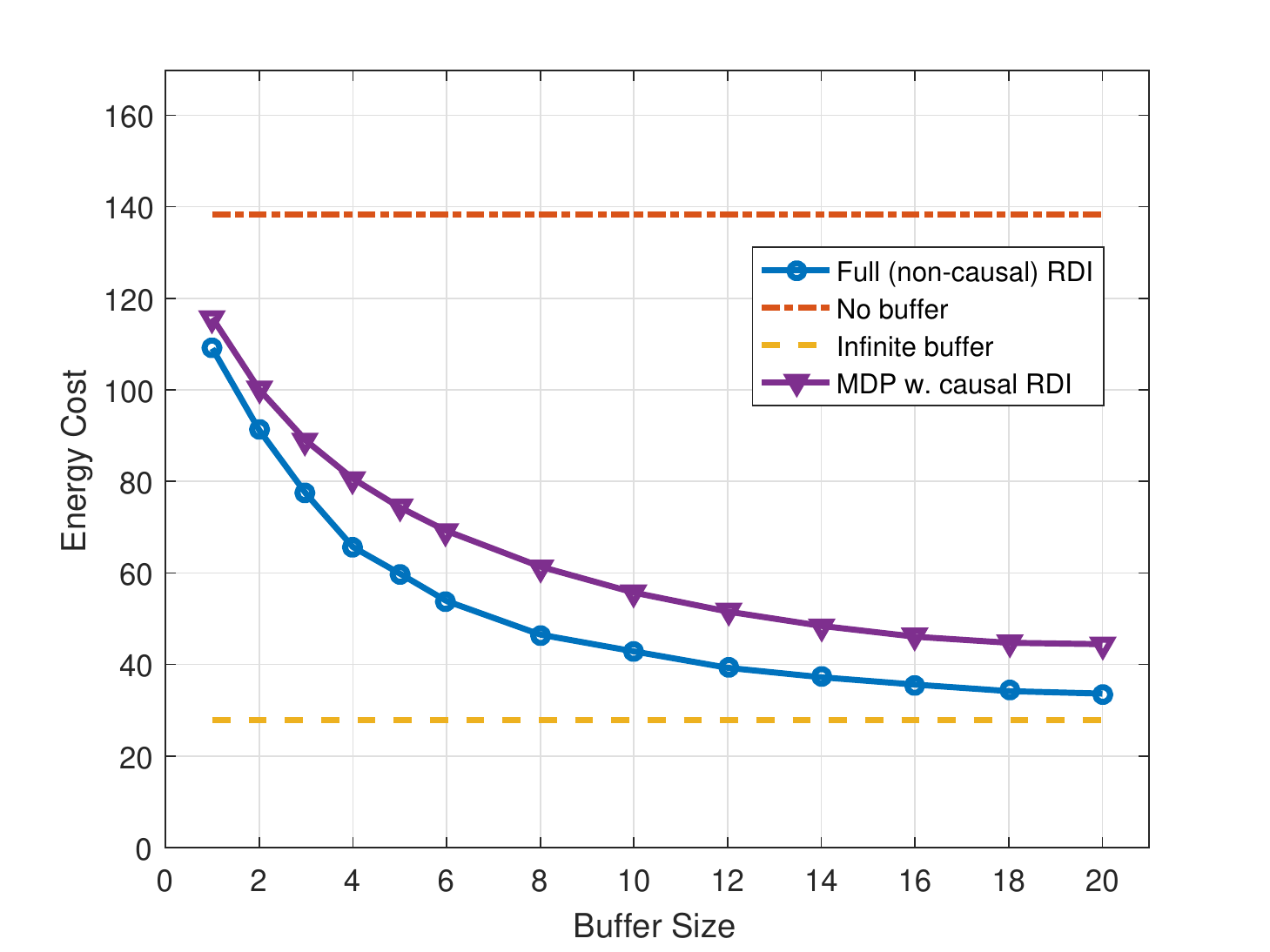}
	\vspace{-0.25cm}
	\caption{Average energy cost decreases as buffer size grows, and the causal MDP method attain similar performance to the non-causal optimum. Here, $x\sim U^d[0,20]$ and $\eta=1.4$.}
	
	\label{sim1}
\end{figure}
%
\begin{figure}[htbp]
	\begin{center}
		\vspace{-0.5cm}
		\includegraphics[width=3in]{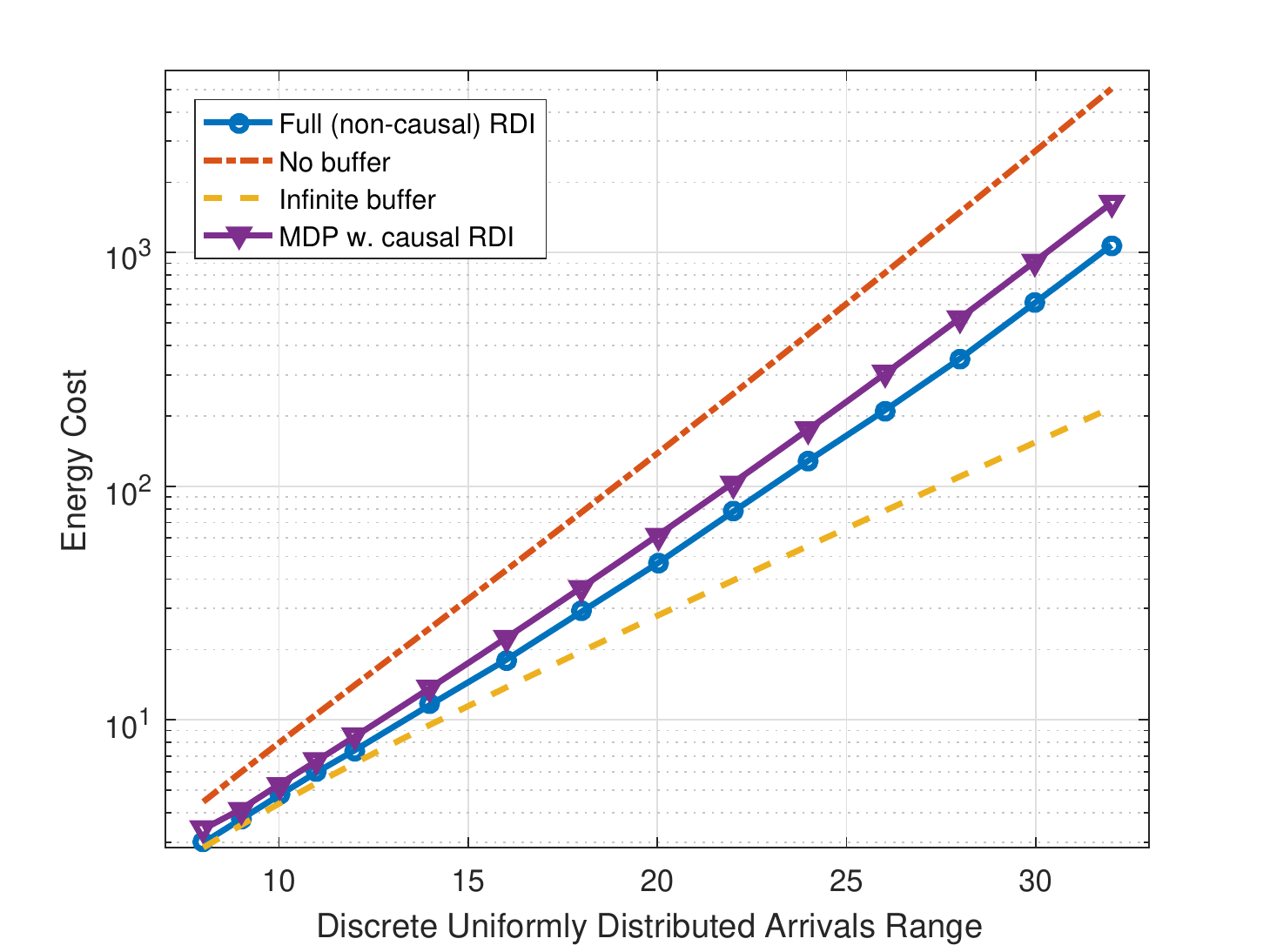}
				\vspace{-0.25cm}
		\caption[Comparision of time complexity]{Average energy cost increases rapidly as data request grows. Similar to the non-causal optimum, causal MDP policy brings significant improvement. Here, buffer size $B=8$ in content items and $\eta=1.4$.}
		\label{sim3}%
	\end{center}%
\end{figure}%
%
\begin{figure}[htbp]
	\begin{center}
		\vspace{-0.8cm}
		\includegraphics[width=3in]{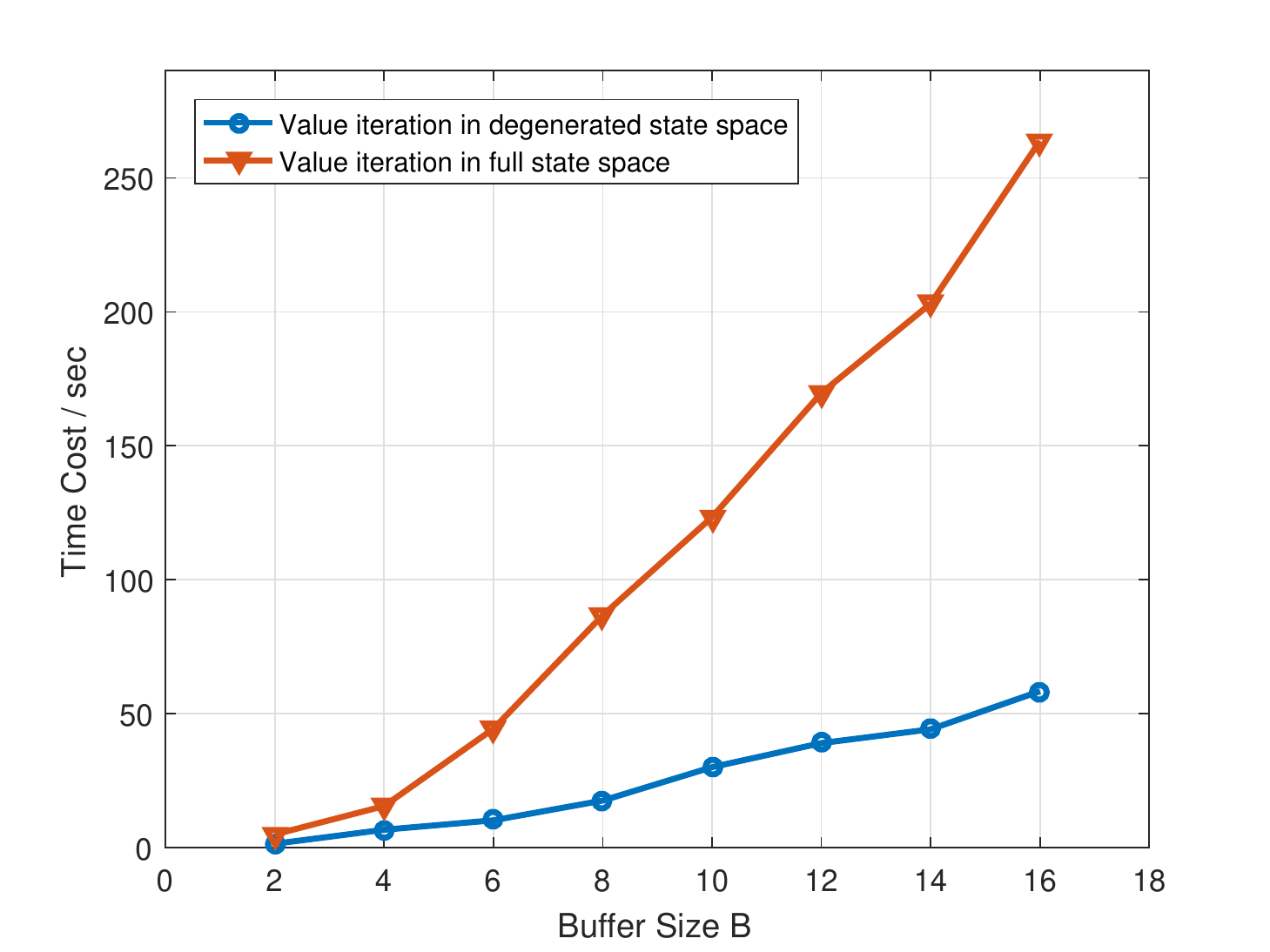}
				\vspace{-0.25cm}
		\caption[Comparision of time complexity]{Value iteration in degenerated space $\mathfrak{B}$ with the FAST algorithm significantly saves time compared with that in $\mathcal{S}$. Here, the user request is of size $X=1.5B$ where $B$ denotes the buffer size with $\eta=1.4$}
		\label{sim6}
	\end{center}
\end{figure} 
%
We first assume that data request yields discrete uniform distribution on all integers in $[0,20]$, and buffer size varies. $\eta$ takes typical values $1.4$ and $2$ respectively. Fig. \ref{sim1} shows that the causal policy designed by MDP compromises little compared with the non-causal optimum with full RDI. The curve \textit{No buffer} corresponds to real-time transmission, and \textit{Infinite buffer} shows the energy cost for stationary transmission at the average rate. The two curves together give the upper and lower bounds of energy cost.


We then fix the buffer size and study the tendency of energy cost as data request pattern changes. Assume that data request always yields discrete uniform distribution on all integers in $[0,X]$. As $X$ grows, Fig. \ref{sim3} shows that energy cost grows exponentially. The optimal MDP policy still brings significant improvement, and its performance synchronically grows with the non-causal full-RDI method.

As the simulations are done with identical parameter settings, where $X=1.5B$ always holds and $B$ takes $\{2,4,...,16\}$, Fig. \ref{sim6} shows that value iteration in full state space $\mathcal{S}$ consumes much more time than the method proposed in this paper, which is value iteration in $\mathfrak{B}$ with the FAST algorithm, though the two methods end with the same policy.

\section{Conclusion}

In this paper, we introduced Markov decision processes to the point-to-point proactive caching problem in wireless communications. Though it is possible to directly apply conventional MDP algorithms to design optimal JPC policies, we largely adapted MDP model for the specified problem setting. We revealed and exploited benefits from a special structure, which we proposed as generalized monotonicity. The algorithm designed based on the very structure has significantly accelerated the conventional MDP algorithm in this problem. Additionally, the energy performance of the attained causal policy is comparatively satisfactory, with regard to the globally optimal non-causal policy. Future works may continue to analytically study the improvement of energy consumption, and generalize conclusions in this paper to the scenario of multicasting.

\bibliographystyle{IEEEtran}
\bibliography{./references}

\end{document}